\documentclass[%
 aip,
cp,  
 amsmath,amssymb,
 reprint,%
]{revtex4-2}

\usepackage{xcolor}
\usepackage{bbm}
\usepackage{graphicx}
\usepackage{dcolumn}
\usepackage{bm}

\usepackage[utf8]{inputenc}
\usepackage[T1]{fontenc}
\usepackage{mathptmx}

\newcommand{\beq}{\begin{eqnarray}}
\newcommand{\eeq}{\end{eqnarray}}
\newcommand{\ket}[1]{\ensuremath{| #1 \rangle}}
\newcommand{\bra}[1]{\ensuremath{\langle #1 |}}

\begin{document}

\title{A Quantum Information Theoretic View On A Deep Quantum Neural Network}

\author{Beatrix C. Hiesmayr} 
 \email[Corresponding author: ]{Beatrix.Hiesmayr@univie.ac.at}
\affiliation{University of Vienna, Faculty of Physics, W\"ahringerstrasse 1, 1090 Vienna (Austria).
}



\begin{abstract}
We discuss a quantum version of an artificial deep neural network where the role of neurons is taken over by qubits and the role of weights is played by unitaries. The role of the non-linear activation function is taken over by subsequently tracing out layers (qubits) of the network. We study two examples and discuss the learning from a quantum information theoretic point of view. In detail, we show that the lower bound of the Heisenberg uncertainty relations is defining the change of the gradient descent in the learning process. We raise the question if the limit by Nature to two non-commuting observables, quantified in the Heisenberg uncertainty relations, is ruling the optimization of the quantum deep neural network. We find a negative answer.
\end{abstract}

\maketitle

\section{\label{sec:introduction} Introduction} 
\begin{figure}
\includegraphics[width=0.85\textwidth,keepaspectratio=true]{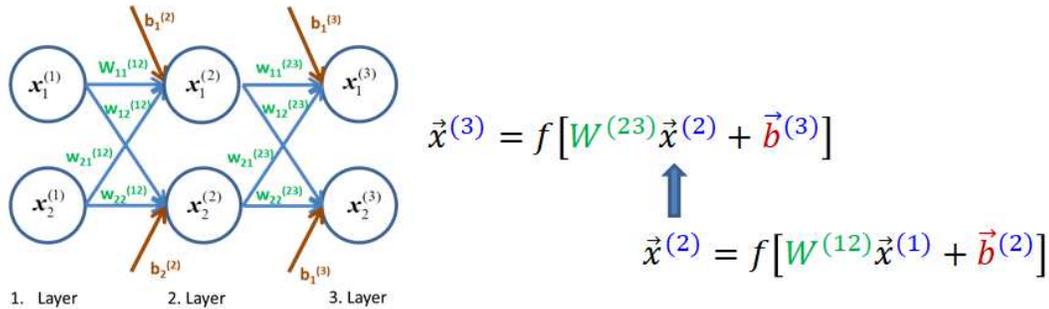}
\caption{\label{fig:cann} This figure sketches how the input vector $\vec{x}^{(1)}$ is processed by feed forward propagation in an artificial deep neural network. The bias and weights are applied to a non-linear function $f$ which defines the input for the next layer and so on.}
\end{figure}

Machine Learning, supervised, unsupervised, reinforcement or GANs (Generative Adversarial Networks), has shown in the recent years impressive successes (e.g. Ref.~\cite{Millinger} for an application to photovoltaic systems or Ref.~\cite{LHCb} for utilizing machine learning for identification of particles or Ref.~\cite{Waltenberger} utilizing deep networks for automatic cleaning of data). Here we want to raise the question if we can do better with quantum systems. In general there are several approaches and claims, but no clear candidate. We focus on quantum artificial neural networks that utilize qubits as perceptrons.  For a general overview over the current perspective of quantum algorithms on a quantum computer in the  noisy intermediate-scale quantum (NISQ) era the reader may e.g. be referred to Ref.~\cite{LeymannBarzen}.

Classical neural networks started their success story once hidden layers were introduced in addition to a non-linear activation function. The very working of classical neural networks is shown in Fig.~\ref{fig:cann}. Weights $w_{k,l}$ and bias $b_{i}$ at different layers are the parameters that have to be learnt by the training pairs provided in the learning process. In addition an activation function $f$ has to be chosen which, as it turned out, has to be non-linear in order to guarantee the universal approximation theorem, i.e. that any function can be  efficiently approximated by this neural network. The learning of the classical network is given by defining a cost function and utilizing backward propagation, which allows to update the weights and bias over a gradient descent such that the cost function optimizes, i.e. the desired output is reached better and better.

We focus on a quantum version of a classical neural network that interchanges each perceptron with a qubit. The weights and maybe bias are realized with different unitary matrices. The main challenge is to introduce an activation function in the quantum version since the quantum theory is manifestly linear and the unitary evolution is reversible and non-dissipative. This is in strong contrast to classical neural networks which have at their heart nonlinear activation functions and a dissipative dynamics. It is generally open \textit{which properties of classical artificial neural networks should be met to call it a meaningful quantum artificial neural network.} But this question goes deeper since it generally asks \textit{what is the difference between classical and quantum information and its processing.}

In this paper we discuss these issues by considering a particular example of a deep quantum artificial neural network.

\section{\label{sec:deepnetwork}A quantum artificial deep neural network}

\begin{figure}
\includegraphics[width=0.7\textwidth,keepaspectratio=true]{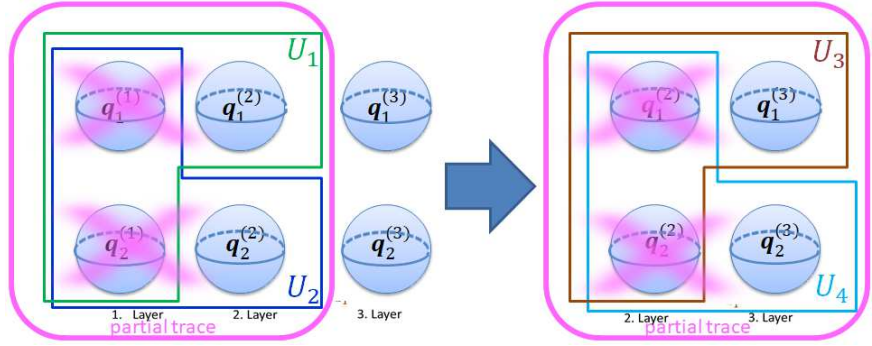}
\caption{\label{fig:qmneuralnetworks} This figure sketches how a two qubit input state is processed by feed forward propagation in an artificial quantum neural network. Two unitaries $U_1$ and $U_2$ are applied followed by a partial trace, the resulting two qubit state is the input for the next layer and so on.}
\end{figure}

A minimal deep quantum artificial neural network is sketched in Fig.~\ref{fig:qmneuralnetworks}. A unitary $U_1$ acts upon two input qubits and the first qubit of the hidden layer, this is followed by a unitary $U_2$ that acts upon the two input qubits and the second qubit of the hidden layer. Obviously, the ordering of these two unitaries is important. Then a partial trace is applied to the input layer -- first two qubits -- resulting in a two qubit state for which the same process is started by two new unitaries $U_3,U_4$ followed by a partial trace over the hidden layer, which is then the two--qubit output state of the quantum neural network. The partial trace may be interpreted as the activation function and the parameters of the four unitary operators $U_1,U_2,U_3,U_4$ as the weights or bias.

For the minimal network of a two qubit input layer ($1.$ layer) and a hidden layer of two qubits ($2.$ layer) and an output layer of two qubits ($3.$ layer) the four unitarities have each the dimension of $8$ which implies $8^2=64$ free parameters and in total $4\cdot 64$ parameters.  As a cost function we will define the fidelity, which is a measure of the ``closeness'' of two quantum states. It expresses the probability that one state will pass a test to identify as the other. It is generally defined by  \begin{eqnarray}
F(\rho,\sigma)&=&F(\sigma,\rho)\;=\;\biggl(Tr(\sqrt{\sqrt{\rho}\sigma\sqrt{\rho}})\biggr)^2\;,\end{eqnarray}
which reduces in the special case of pure states to the overlap of those two states. For qubits the fidelity reduces also to $F(\rho,\sigma)=Tr(\rho\sigma)+2\sqrt{\det(\rho)\det{\sigma}}$. The fidelity takes values $\in[0,1]$ and in the case of $1$ the states can be considered as equivalent. The problems that we will consider will have a desired output state $\Phi_{desired}^z$ that is chosen to be pure. This simplifies the loss function to $C=\langle \Phi_{desired}| \rho_{out} |\Phi_{desired}\rangle$, which is $1$ if the output state of the network $\rho_{out}$ perfectly overlaps with the desired state $\Phi_{desired}$ and else $\in[0,1\}$.

\subsection{Feed Forward Propagation}

The two-qubit output state of the minimal network of ``two-qubits--two-qubits--two-qubits'' is given by
\begin{eqnarray}
\rho_{out}^z&=&\sum_{i,j,k,l=0}^1 \langle ij kl|\otimes \mathbbm{1}_4\; U_4\; U_3\; U_2\; U_1\;\;\rho_{in}^z \otimes|00\rangle\langle00|\otimes|00\rangle\langle00|\;\; U_1^\dagger\; U_2^\dagger\;U_3^\dagger\; U_4^\dagger \;\;|ij kl\rangle\otimes\mathbbm{1}_4\;,
\end{eqnarray}
where $\rho_{in}^z$ is a given input state, the initial states of the hidden and output layers have been chosen to be in the state $|0\rangle$ (w.o.l.g.) and the unitaries only address the subspaces described before.

\subsection{Cost Function Optimization - Backward Propagation}

The cost function for our problem is then defined by
\beq
C(\{\Phi_{desired}^z\}_{z=1}^{N},\{\rho_{in}^z\}_{z=1}^{N},U_1,U_2 ,U_3 ,U_4)&:=&\frac{1}{N}\sum_{z=1}^{N}\;\langle \Phi_{desired}^z| \rho_{out}^z |\Phi_{desired}^z\rangle\;. 
\eeq
In Ref.~\cite{unitaryparameterizationHiesmayr} a composite parametrization was introduced which will allow us to compute the derivative of the unitaries, for the optimization of the network, analytically. For any unitary operation $U$ acting on a Hilbert space $\mathcal{H}=\mathbb{C}^d$ with $d\geq2$ spanned by the orthonormal basis $\{ \ket{1},\ldots,\ket{d} \}$ there exist $d^2$ real values $\lambda_{m,n}$ with $m,n \in \left\{ 1 , \ldots , d \right\}$ and ${\color{red}\lambda_{n,n}} \in \left[0, 2 \pi \right]$ and ${\color{blue}\lambda_{n,m}} \in \left[0, 2 \pi \right]$ for $m < n$ and ${\color{green}\lambda_{m,n}} \in \left[0, \frac{\pi}{2} \right]$ for $m < n$ such that any $U\equiv U_C$ with
\begin{align}
\label{Uc}
U_C=\left[\prod_{m=1}^{d-1} \left(\prod_{n=m+1}^{d} \underbrace{\mbox{exp} \left( i\, {\color{blue}{\lambda_{n,m}}}\;P_n  \right) \mbox{exp} \left( i\, {\color{green}{\lambda_{m,n}}}\;Y_{m,n} \right)}_{:=\Lambda_{m,n}}  \right) \right] \cdot \left[ \prod_{l=1}^{d} \mbox{exp}(i\, {\color{red}{\lambda_{l,l}}}\;P_l )\right] \ .
\end{align}
The sequence of the product is defined by $\prod_{i=1}^{N}A_i=A_1 \cdot A_{2} \cdots A_N$.
Here, the $P_l $ are one-dimensional projectors
$P_l=\ket{l-1}\bra{l-1}$
and $Y_{m,n}$ are the generalized anti-symmetric Pauli-matrices
$Y_{m,n}=-i\ket{m-1} \bra{n-1} + i \ket{n-1} \bra{m-1}$
with $1\leq m < n \leq d$.

The parameter $\lambda_{m,n}$ can be gathered in a $d \times d$ ``\textit{parameterization matrix}''
\begin{align}
\textrm{``\textit{Parameterization matrix}''}\;\equiv\;\left(
  \begin{array}{ccccc}
    {\color{red}{\lambda_{1,1}}} &{\color{green}{\lambda_{1,2}}}&   \cdots &{\color{green}{\lambda_{1,d-1}}}& {\color{green}{\lambda_{1,d}}} \\
    {\color{blue}{\lambda_{2,1}}} &{\color{red}{\lambda_{2,2}}}&   \cdots &{\color{green}{\lambda_{2,d-1}}}&{\color{green}{\lambda_{1,d}}} \\
    \vdots &\vdots & \ddots & \vdots \\
   {\color{blue}{\lambda_{d-1,1}}} &{\color{blue}{\lambda_{d-1,2}}}&   \cdots &{\color{red}{\lambda_{d-1,d-1}}}& {\color{green}{\lambda_{d-1,d}}} \\
    {\color{blue}{\lambda_{d,1}}} &{\color{blue}{\lambda_{d,2}}}&   \cdots &{\color{blue}{\lambda_{d,d-1}}}& {\color{red}{\lambda_{d,d}}}
  \end{array}
\right) \ ,
\end{align}
where the diagonal entries ${\color{red}{\lambda_{n,n}}}$ represent global phase transformations, the upper right entries ${\color{green}{\lambda_{m,n}}}$ are related to rotations in the subspaces spanned by $|n\rangle$ and $|m\rangle$, while the lower left entries ${\color{blue}{\lambda_{n,m}}}$ are relative phases in these subspaces (with respect to the basis $\{\ket{0},\ldots,\ket{d-1}\}$). Note that for optimization one does not need to restrict the parameter $\lambda_{n,m}$ to the intervals given above.

Now we want to change the unitaries of the neural network in order to maximize the cost function $C$ and this for each parameter $\lambda_{xy}$, i.e. we can consider a Taylor expansion:
\beq
U(\lambda_{x,y}+\varepsilon)&\doteq& U(\lambda_{x,y})+\varepsilon \frac{\partial U(\lambda_{x,y})}{\partial \lambda_{x,y}}+O(\varepsilon^2)\nonumber\\
&=&U(\lambda_{x,y})+i\;  \varepsilon U(\lambda_{x,y})\; \Tilde{Y}_{x,y}+O(\varepsilon^2)
\eeq
with
\beq
\Tilde{Y}_{xy}&=& \left\lbrace\begin{array}{l} U^\dagger_{x,y}\;Y_{x,y} U_{x,y}\qquad\textrm{for}\quad x<y\\
P_x\qquad\qquad\qquad\textrm{for}\quad x=y\\
U^\dagger_{x,y}\;P_{x} U_{x,y}\qquad\textrm{for}\quad x>y\;,\end{array}\right.
\eeq
and
\beq
U_{x,y}=\left\lbrace\begin{array}{l}\left[\Pi_{n=y+1}^{d-1} \Lambda_{x,n}\right]\left[\Pi_{m=x+1}^{d-2} \Pi_{n=m+1}^{d-1} \Lambda_{m,n}\right]\;\Pi_{l=x}^{d-1} e^{i P_l \lambda_{l,l}}\qquad\textrm{for}\quad x<y\\
\left[\Pi_{n=x}^{d-1} \Lambda_{y,n}\right]\left[\Pi_{m=y+1}^{d-2} \Pi_{n=m+1}^{d-1} \Lambda_{m,n}\right]\;\Pi_{l=y}^{d-1} e^{i P_l \lambda_{l,l}}\qquad\textrm{for}\quad x>y\end{array}\right.\;
\eeq
where we have used the results of Ref.~\cite{unitaryparameterizationHiesmayr}. Note that $\Tilde{Y}$ are hermitian, thus the unitarity condition holds for every order in the expansion. Moreover, note that  $\Tilde{Y}$ depends for $x\not= y$ on all other $\lambda$ parameters except $\lambda_{x,y}$.

Thus the parameters of the unitaries are changed by $\lambda_{x,y}\longrightarrow \lambda_{x,y}+\;\delta \lambda_{x,y}$ which is given for the first perceptron by the term
\beq
\delta \lambda^{(1)}_{x,y}\approx  \varepsilon\frac{\partial C^z}{\partial \lambda^{(1)}_{x,y}}\approx \varepsilon \sum_{ij=0}^1 \langle ij \Phi_{desired}^z|\; U_2 U_1\;\;{\color{blue} i\; [\Tilde{Y}^{(1)}_{x,y},\tilde{\rho}^z]}\;\; U_1^\dagger U_2^\dagger\;\; |ij\Phi_{desired}^z\rangle\;,
\eeq
and for the second perceptron by the term
\beq
\delta \lambda^{(2)}_{x,y}\approx \varepsilon\frac{\partial C^z}{\partial \lambda^{(2)}_{x,y}}\approx  \varepsilon \sum_{ij=0}^1 \langle ij \Phi_{desired}^z|\; U_2\;\;{\color{blue} i\; [\Tilde{Y}^{(2)}_{x,y},U_1\tilde{\rho}^z U_1^\dagger]}\;\; U_2^\dagger\;\; |ij\Phi_{desired}^z\rangle\;,
\eeq
and so on. Here $\varepsilon$ can be chosen arbitrarily and in principle different for each unitary and plays the role of a learning parameter in a classical network, i.e. chosen too low the cost function will only increase slowly but chosen too high we may miss the optimum. It is a hyper parameter in the learning process. We chose it for all four unitaries the same in our applications.

Let us emphasize here that each equation forms a Heisenberg uncertainty, in the so called Robertson version~\cite{Robertson}, i.e.
\beq\label{HeisiEr}
(\Delta \Tilde{Y}_{x,y})_\Psi \Delta (\tilde{\rho})_\Psi\;\geq \frac{1}{2}\left|i\; \langle \Psi| \; \left[\Tilde{Y}_{x,y},\tilde{\rho}\right]\; |\Psi\rangle\right|\;
\eeq
where the $\Delta$ is the standard deviation of the operator with respect to $\Psi$. Clearly, there are only two ways how the lower bound on a Heisenberg uncertainty can vanish, either the two observables are commuting or the state $\Psi$ has a spectrum of zero. The first way is the general foundational limit provided by Nature, if two observables are not commuting, for instance the famous position operator $\hat{x}$ and momentum operator $\hat{p}$, we have $[\hat{x},\hat{p}]=i\hbar\mathbbm{1}$ and therefore, for all possible states $\Psi$ the lower bound is $\frac{\hbar}{2}$. Differently stated, there exists no states for which the standard deviation of the position and momentum can be smaller than this value.

On the other hand if we consider e.g. Pauli operators $\sigma_i$, then the commutator is $[\sigma_i,\sigma_j]=2 i \varepsilon_{ijk} \sigma_k$ and thus the lower bound  gives
\beq
|\langle \Psi|\sigma_k|\Psi\rangle|
\eeq
which may be a non-zero value for a general $\Psi$, but choosing an appropriate $\Psi$ it may still vanish though the Pauli operators do not commute. This property of the Robertson version of the Heisenberg uncertainty relation was criticized and an entropic version was found overcoming this issue, which we discuss in the conclusions. The question we want to discuss first is if in the optimization process of the quantum net, these fundamental limits are utilized.

\section{Examples and Results}

Here we present two different examples by increasing the complexity of the general problem.

\subsection{Example A: Learning A Single Unitary}

Let us start with a simple example, namely learning a particular unitary $V$, that was first considered in Ref.~\cite{Beer}. In Ref~\cite{WilkinsonHartmann} this quantum neural network was applied to the real data of the Iris flower and the performance of this quantum network was compared with other networks. The ground truth is then given by choosing $z$ arbitrary states $\rho^z$ and computing the desired output by $\rho_{desired}^z=V \rho_{in}^z$. The goal is that the network learns this unitary (generally only $16$ parameters) by optimizing the $4\cdot 64$ parameters of the network by utilizing the cost function. One can fix $\varepsilon$, which may be interpreted as a learning parameter, in each round but we optimize $\varepsilon$ by taking the maximum of the cost function for the computed corrections of all $4$ unitaries. We chose randomly $100$ pairs and used $10$ for the optimization and $90$ for the validation. The software for implementation was Mathematica Wolfram.

\begin{figure}
\includegraphics[width=0.5\textwidth,keepaspectratio=true]{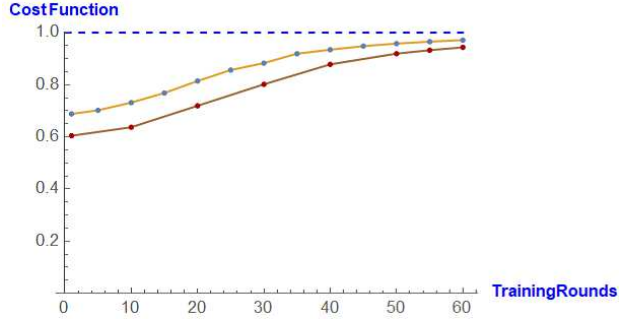}
\caption{The curves show the (orange-blue) cost function ($10$ pairs) and the (red-brown) validation function ($90$ pairs) in dependence of the training rounds for learning a single unitary. Actually, the cost function started with a value of $0.25$, we show here only the optimization after the value $0.6$ was reached.}\label{fig:singleunitarycostfunction}
\end{figure}

In Fig.~\ref{fig:singleunitarycostfunction} we plotted the cost function and the validation function for different training rounds. At each training round the cost function is plotted as a function of $\varepsilon$ and its maximum is taken. As can be seen the curves are monotonically increasing at each round but the convergence is slow.

\begin{figure}
\includegraphics[width=0.7\textwidth,keepaspectratio=true]{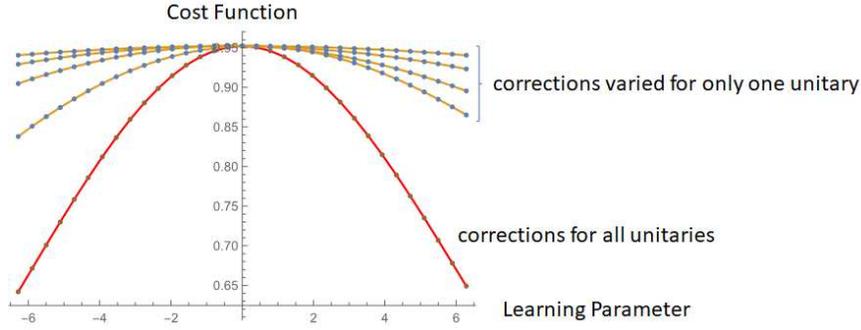}
\caption{The curves show the cost function $C$ varied with the learning parameter $\varepsilon$ for a typical step in the optimization process. If all corrections to the four unitaries are included, the maximum is quite pronounced and with that the choice of the optimal $\varepsilon$ for the problem at each round. Varying only one unitary the maximum is less pronounced, i.e. the problem is less constrained. Further note that the problem is not symmetric due to the composite parameterization of the unitaries.}\label{fig:learningparameter}
\end{figure}

A typical update for the $\lambda_{x,y}$ looks like  (example for $U_1$, round $55$)
\beq
\tiny{\left(
\begin{array}{cccccccc}
 0.0183027 & -0.0139361 & 0.000587036 & -0.00177987 & 0.00189422 & -0.00494903 & 0.00476928 & 0.00521779 \\
 -0.0110207 & 0 & -0.00441354 & 0.00138528 & 0.00458157 & 0.000161181 & -0.000201973 & -0.0110672 \\
 -0.0000499063 & -0.000143331 & -0.00882844 & 0.00517684 & -0.00501275 & -0.0026375 & 0.00608451 & 0.00746686 \\
 -0.000666153 & -0.000108606 & -0.00185629 & 0 & 0.00584462 & -0.00359862 & 0.0000593791 & 0.000123862 \\
 -0.00363472 & 0.00757916 & -0.00248305 & -0.00310511 & -0.0106122 & 0.00191896 & 0.000186693 & -0.000654047 \\
 0.00162855 & -0.0133182 & 0.000649396 & 0.00231792 & 0.001892 & 0 & -0.00185319 & 0.00141976 \\
 0.017559 & 0.0012003 & -0.00531224 & -0.000393857 & -0.00213533 & 0.000525167 & 0.00113794 & 0.00242888 \\
 -0.0079773 & -0.0082346 & -0.00808296 & -0.00738433 & -0.00711041 & -0.000057534 & 0.00043083 & 0 \\
\end{array}
\right)}
\eeq
or for (example for $U_1$, round $60$)
\beq
\tiny{\left(
\begin{array}{cccccccc}
 -0.00310993 & -0.00480453 & -0.000210951 & -0.000935706 & 0.00180615 & -0.00740156 & 0.00205214 & -0.00979687 \\
 0.00169218 & 0 & 0.000343261 & -0.000831133 & 0.000124329 & -0.00183596 & 0.00103077 & -0.00848936 \\
 -0.000158018 & -0.000150325 & 0.000334237 & -0.00343378 & 0.00233581 & -0.00165355 & 0.00201545 & 0.00609294 \\
 -0.000131329 & 0.000143296 & 0.00110842 & 0 & -0.00292121 & 0.00247819 & 0.00484749 & -0.000320837 \\
 -0.00370207 & 0.00683878 & 0.000496848 & 0.00216435 & 0.00432188 & 0.000335835 & 0.00236745 & -0.00040123 \\
 0.00308359 & -0.0109918 & 0.00117493 & 0.000951308 & -0.000982794 & 0 & -0.00144647 & 0.00150626 \\
 -0.00397976 & 0.00264269 & -0.00276574 & 0.00147698 & 0.00389812 & 0.0000669185 & -0.00154618 & 0.00125958 \\
 0.00116858 & 0.000667512 & 0.000810526 & -0.0028471 & -0.00113776 & -0.0000591576 & 0.00000439 & 0 \\
\end{array}
\right)}
\eeq
which is the sum of all training states defining the lower bound in the Heisenberg relation and does not vanish. The zeros are due to the fact that we have chosen the hidden layer and output layer states to be $|0\rangle$. In Fig.~\ref{fig:learningparameter} we show how the cost function typically changes with the learning parameter $\varepsilon$. It is quite constrained if all four unitaries are included, but for a single one the parameter space is quite flat. This suggests that the interplay of all four unitaries is relevant for the problem, but the constraint due to each single unitary does not do the job.

Even  though we are close to the maximum cost function value it seems that the derivatives do not vanish. To see if this is due the average over in this case $10$ pairs, we picked one out and optimized it to a cost function value of $0.999003$, the corrections terms are still of same order as above. This suggests that the neural network does not optimize the uncertainty relation but the parameters of the unitaries which are not unique. Let us choose now a non-trivial problem.

\subsection{Example B: A State Learning Its Own Quantum Properties?!}

Now we create pairs such that the desired output state encodes the quantum properties, i.e. its purity $\mathsf{P}(\rho)=\frac{4}{3}(Tr(\rho^2)-\frac{1}{4})$ and its entanglement property. For that purpose we choose the concurrence~\cite{HillWootters}, a computable measure for bipartite qubits. The concurrence $\mathsf{Con}(\rho)$ is defined as maximal between zero and the maximal eigenvalue minus the other three eigenvalues of the quantity $\sqrt{\sqrt{\rho}\sigma_2\otimes\sigma_2\rho^*\sigma_2\otimes\sigma_2\sqrt{\rho}}$ with $\sigma_2$ being the $y$-Pauli matrix. For  pure states it simplifies to $\mathsf{Con}(\psi)\;=\;|\langle \psi|\;\sigma_y\otimes\sigma_y|\psi^*\rangle|$.

Our desired output states are chosen to be
 \beq
 |\Phi^{z}_{desired}\rangle&=& \sqrt{\frac{\mathsf{Con}(\psi_{in}^z)}{2}}\; |00\rangle+\sqrt{1-\frac{\mathsf{Con}(\psi_{in}^z)+\mathsf{P}(\psi_{in}^z)}{2}} \frac{1}{\sqrt{2}}(|01\rangle+|10\rangle)+\sqrt{\frac{\mathsf{P}(\psi_{in}^z)}{2}}\;|11\rangle\;.
 \eeq
This means that each pair is again connected by a unitary (if we assume only pure input states), $ |\Phi^{z}_{\mathsf{desired}}\rangle=U_{\psi^{z}_{in}} |\psi^{z}_{in}\rangle$, but it is chosen according to the quantum properties of the input states. Thus the net needs to learn a set of unitaries defined by the quantum properties (entanglement\&purity) of the arbitrary input state.  Consequently, the question is whether the neural net processes also the properties of the state itself or if only the information of the training pair is exploited as it would be the case in a classical neural network.

We tried different sets for the training and here we discuss the result for $70$ pairs for the training. The convergence is even slower than for the problem of single unitaries. We find typically a cost function value of $0.91$ with a standard deviation of $0.08$. If we use for the validation only $30$ pairs the cost function value was found to be higher, i.e. for our set $0.93$. This shows a high statistic fluctuation with the randomly chosen set, meaning the general problem is not (yet) fully learnt. As a further test we can  interchange output with the input, this gave a cost function of $0.30$ with a standard deviation of $0.26$. We also tried random inputs, which gave in general very low cost function values. Consequently, the net is indeed learning some features of the training set, which also applies to an arbitrary set.

\begin{figure}
(a)\includegraphics[width=0.9\textwidth,keepaspectratio=true]{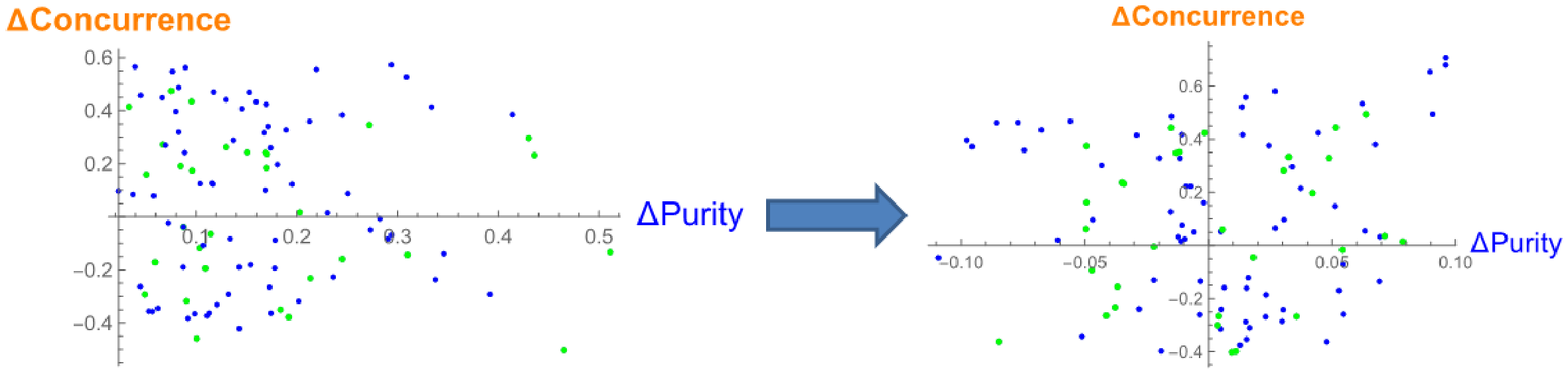}
(b)\includegraphics[width=0.9\textwidth,keepaspectratio=true]{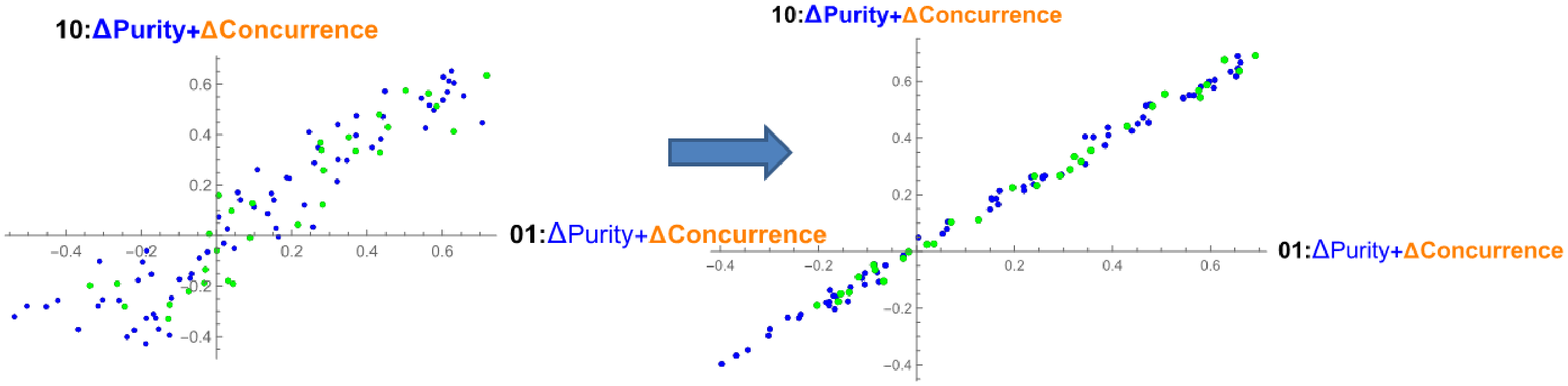}
\caption{The pictures show the differences between the output of the net and the desired output for an early stage of optimization and at the last round for a (noise free) measurement of (a) $00$ and $11$ and (b) $01$ and $10$. Blue dots represent the $70$ training pairs and green dots the $30$ validation pairs.}\label{fig:conpurity}
\end{figure}

In Fig.~\ref{fig:conpurity} we visualize how well the quantum properties are learnt per se. The first graphs correspond to an early time in the optimization process, here the cost function gave the value $0.84$ with a standard deviation of $0.09$. We see that in the optimization process the net learns e.g. the symmetry between the $|01\rangle$ and $|10\rangle$ states (Fig.~\ref{fig:conpurity}(b)), but the range of the errors does not get smaller when compared to a later stage in the optimization. From Fig.~\ref{fig:conpurity}(a) we can deduce that the error in purity is significantly reduced (having only pure states in the training), but the system also predicts values greater than $1$, which is of course unphysical. This could be compensated by adding a Lagrange multiplier to the cost function. In general we observe that the training and validation pairs distribute quite similarly. The range in the error of the concurrence (Fig.~\ref{fig:conpurity}(a)), however, is not reduced.

\begin{figure}
\includegraphics[width=0.5\textwidth,keepaspectratio=true]{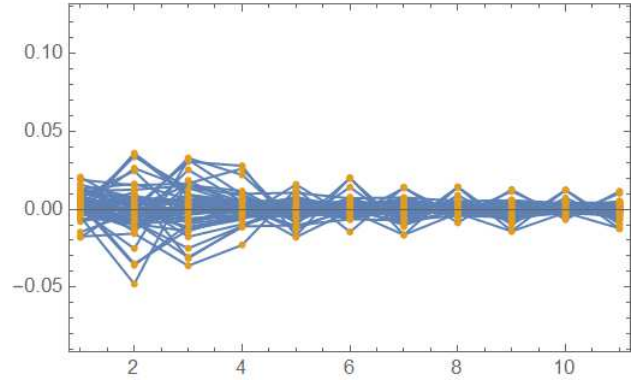}
\caption{This figure shows the $64$ corrections of each $\lambda_{x,y}$ forming one unitary for different optimization rounds. After some optimization steps we see oscillations but no longer differences in the sizes of the corrections in average.}\label{fig:lambdaparameters}
\end{figure}

\begin{figure}
(a)\includegraphics[width=0.47\textwidth,keepaspectratio=true]{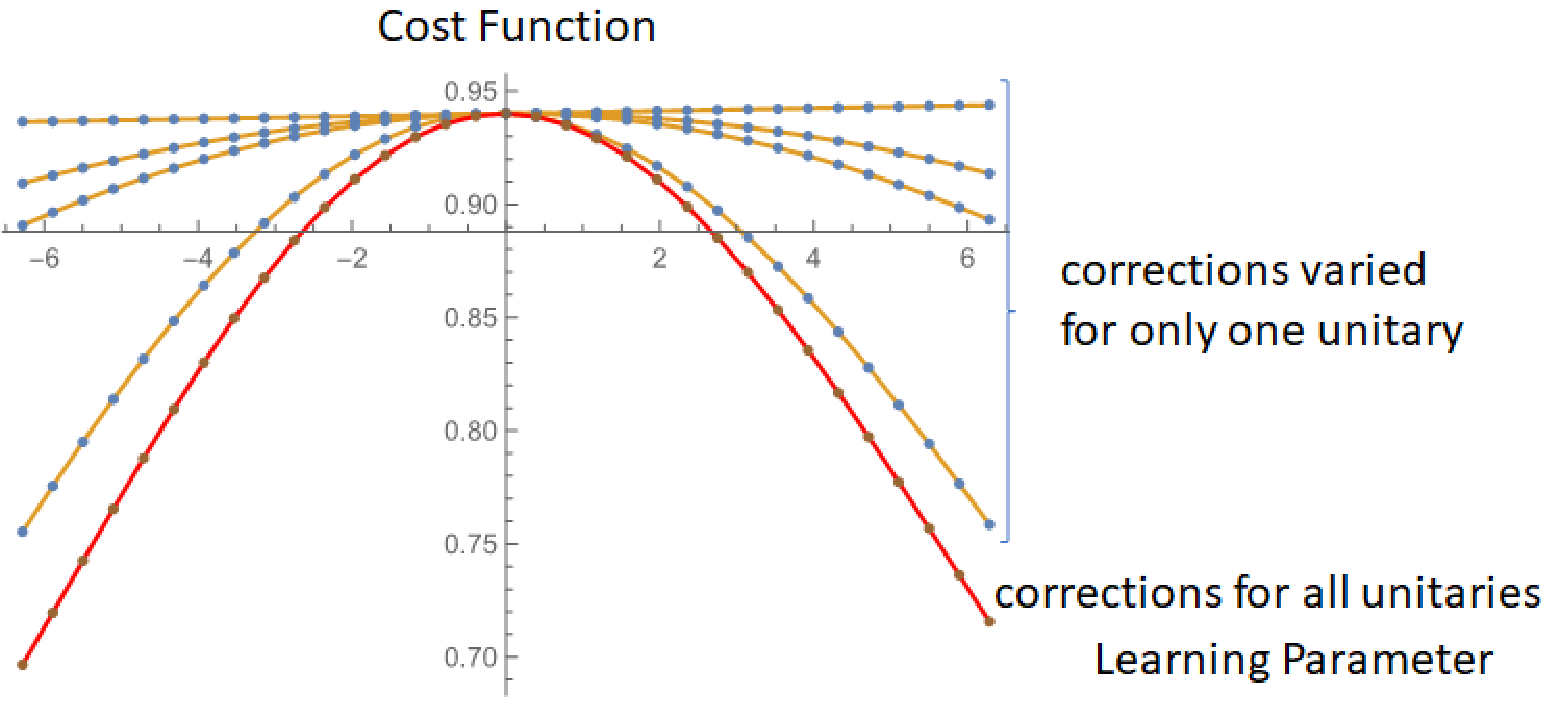}
(b)\includegraphics[width=0.47\textwidth,keepaspectratio=true]{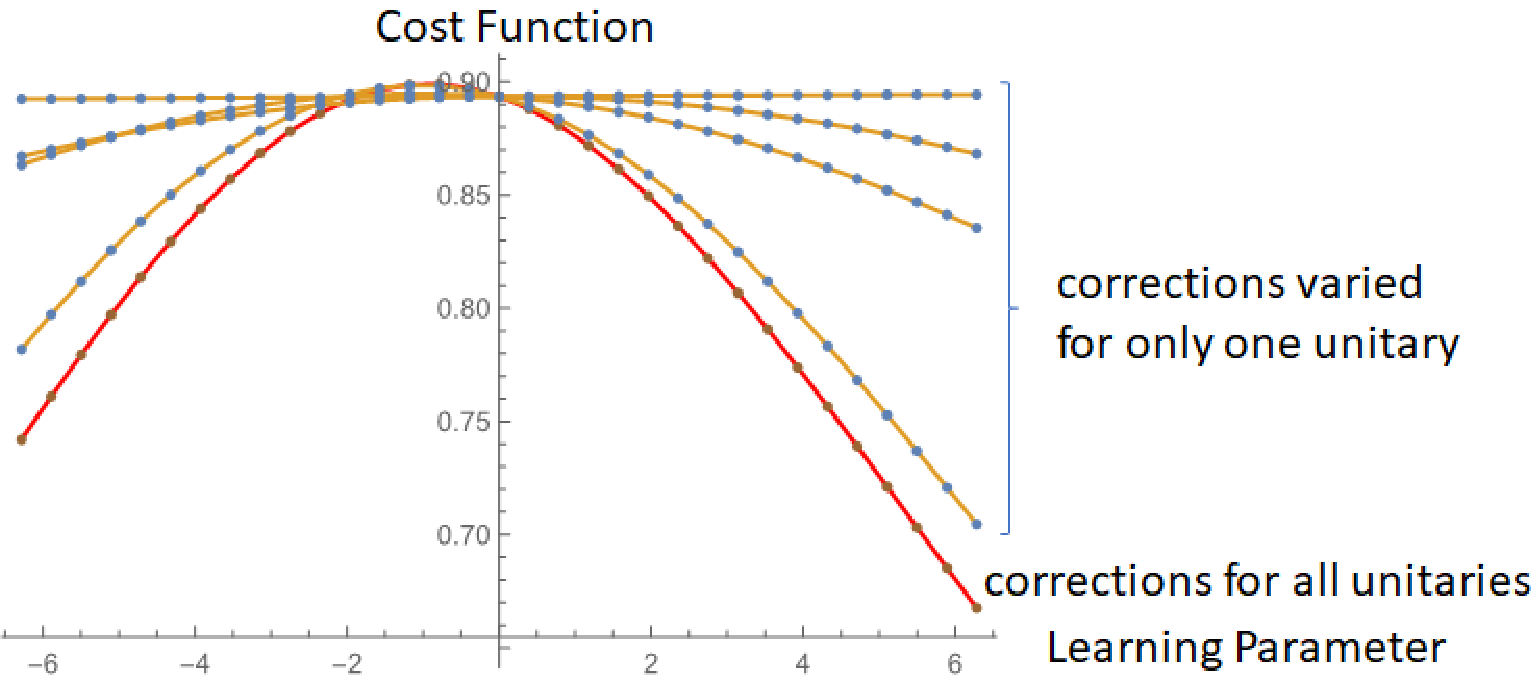}
\caption{The curves show the cost function $C$ varied with the learning parameter $\varepsilon$ for a typical step in the optimization for (a) the training ($30$ states) and (b) the validation set ($70$ states). If all corrections to the four unitaries are included, the maximum is quite pronounced and with that the choice of the optimal $\varepsilon$ for the problem. Varying only one unitary the maximum is less pronounced, i.e. the problem is less constrained except for one of the four unitaries. In contrast to the training set (a) the validation set (b) does not give the maximum in the parameter space and a considerably lower value of the cost function as expected. However, the parameter--space for the learning parameter $\varepsilon$ is quite similar to the training set.}\label{fig:learningparameter2}
\end{figure}

The correction terms obtained by back propagation are always of similar size similar to the trivial example discussed in the previous section and have been visualized in Fig.~\ref{fig:lambdaparameters}. The dependence on the learning parameter $\varepsilon$ is depicted in Fig.~\ref{fig:learningparameter2}. In conclusion, the net learns partial properties of the set but the cost function does converge slowly. In the next section we discuss if the learning exploits the limits by the Heisenberg uncertainty relation.

\section{Conclusion \& Discussion}

In this contribution we have analysed a minimal deep quantum neural network, i.e. a net taking as an input a two--qubit state and producing a two--qubit state as an output, with one hidden layer of two qubits in between. For that we performed two case studies. In the first case, each training pair is connected by one particular randomly chosen unitary matrix. In the second example each pair is connected by a unitary that allows to deduce from the output states the concurrence, a measure of entanglement, and the purity of the input state. Hence here we ask whether the net also learns those implicit properties of the input state, which is obviously classically impossible. In both cases at each round of optimization the cost function was always strictly increasing but typically not by a huge amount. Consequently, some learning of the net has always been observed.

The unitaries involved in the net have been parameterized in a composite way, which allows a quantum information theoretic view into the working of the net. In particular it shows that the corrections to such parameterized unitaries are of the form of a Heisenberg uncertainty relation, Eq.~(\ref{HeisiEr}). One striking feature of the Quantum Nature of our world is that two non-commuting observables lead in general to a universal limit by Nature on the measurement outcomes. The most famous example is the uncertainty in momentum and position, $(\Delta x)_\psi (\Delta p)_\psi\geq \frac{\hbar}{2}$. The fact that the lower bound, the universal limit by Nature, is independent of the state is a special property of those two observables. In general one obtains the quantity $\frac{1}{2}\left|\langle \psi|[\hat{A},\hat{B}]\right|\psi\rangle|$. This is the Robertson form~\cite{Robertson} of the Heisenberg uncertainty relation and was criticized, because by choosing an appropriate state  $\psi$ it can vanish even if the two observables $\hat{A},\hat{B}$ do not commute.

Furthermore, it was shown that there exists an information theoretic formulation  of uncertainty principle~\cite{Deutsch}, which does not suffer from this problem of state dependence. This puts a limit on the extent to which the two observables can be simultaneously peaked. This entropic uncertainty relation of two non-degenerate observables is given by (introduced by D. Deutsch~\cite{Deutsch}, improved in Ref.~\cite{Kraus} and proven in Ref.~\cite{Uffink})
\begin{eqnarray}\label{EQUI}
S(\hat{A})+S(\hat{B})&\geq& 
- \log_d\left(\max_{i,j}\{\left\vert\langle\chi_a^i|\chi_b^j\rangle\right\vert^2\}\right)
\end{eqnarray}
where  \begin{eqnarray}
S(\hat{A})=-\sum_a p_a\log_d p_a
\end{eqnarray}
is the entropy  for e.g. a certain prepared pure state $\psi$ and the $p_a$ is the probability associated with the measurement of  outcome $a$ of $\hat{A}_a$ for $\psi$,  hence $p_a=|\langle \chi_a|\psi\rangle|^2$. Thus in general there is a universal limit to any two observables if they are non-commuting.

Coming back to our quantum neural network. We observed that those lower bounds never vanish, not even for a single generator $\tilde{Y}$. From that we can conclude that the net does not optimize the unitaries involved such that all or some commutators vanish. Consequently, we can conjecture that  the universal limit is not exploited in the optimization. Rather the fact that the parameters are oscillating shows the similarity to classical neural networks optimization. From that we infer that the optimization process does not exploit a particular quantum phenomenon.

In summary, those preliminary results have to be taken with care since we only used a minimal version of a net, e.g. no deeper nets, only one example of a gradient descent-based optimization and a limited set of problems. Moreover, there are several more techniques that could be applied to optimize the learning process. Utilizing a gradient descent-based optimization our findings are also strongly correlated to other works, e.g. Refs.~\cite{GD1,GD2,GD3}, discussing e.g. barren plateau landscapes and how to avoid them. Further detailed studies are necessary to confirm these findings. However, for this minimal setting discussed here Heisenberg's uncertainty relation is not a guiding principle.

\begin{acknowledgments}
BCH thanks the organizers of the workshop ``\textit{International Workshop on Machine Learning and Quantum Computing, Applications in Medicine and Physics (WMLQ2022)}'' for putting together an inspiring and at the top of the knowledge programme and a vivid environment for discussions. BCH also acknowledges gratefully that this research was funded in whole, or in part, by the Austrian Science Fund (FWF) project P36102.
\end{acknowledgments}

\nocite{*}
\bibliography{aipsamp}

\end{document}